\documentclass[final,5p,times,twocolumn]{elsarticle} 
\usepackage{hyperref}
\usepackage{subcaption}
\journal{}

\begin{document}
\begin{frontmatter}
\title{Next Generation Multi-element monolithic Germanium detectors for Spectroscopy: First integration at ESRF facility }

\author[SOLEIL]{N. Goyal\corref{mycorrespondingauthor}}
\address[SOLEIL]{Synchrotron SOLEIL, L'Orme des Merisiers Départementale 128, Saint-Aubin, France}
\cortext[mycorrespondingauthor]{Corresponding author}
\ead{nishu.goyal@synchrotron-soleil.fr}
\author[XFEL]{S. Aplin}
\address[XFEL]{ European XFEL, Schenefeld, Germany}
\author[INFN]{A. Balerna}
\address[INFN]{INFN, Frascati National Laboratory, Frascati (RM), Italy}
\author[MAX]{P. Bell}
\address[MAX]{MAX IV Laboratory, Lund, Sweden}
\author[ALBA]{J. Casas}
\address[ALBA]{ALBA – CELLS Synchrotron, Cerdanyola del Vallès, Spain}
\author[MAX]{M. Cascella}
\author[DLS]{S. Chatterji}
\address[DLS]{Diamond Light Source Ltd, Oxfordshire, United Kingdom}
\author[ESRF]{C. Cohen}
\address[ESRF]{ European Synchrotron Radiation Facility ESRF, Grenoble, France}
\author[ESRF]{E. Collet}
\author[ESRF]{P. Fajardo}
\author[DLS]{E.N. Gimenez}
\author[DESY]{H. Graafsma}
\address[DESY]{German Electron Synchrotron DESY, Hamburg, Germany}
\author[DESY]{H. Hiresmann}
\author[SOLEIL]{F.J. Iguaz}
\author[MAX]{K. Klementiv}
\author[sol]{K. Kolodjiez}
\address[sol]{ National Synchrotron Radiation Centre SOLARIS, Kraków, Poland }
\author[SOLEIL,LAPP]{L. Manzanillas}
\address[LAPP]{Laboratoire d'Annecy de Physique des Particules, LAPP - CNRS/IN2P3, Annecy, France}
\author[ESRF]{T. Martin}
\author[ELET,tri,swe]{R. H. Menk}
\address[ELET]{ Elettra Sincrotrone Trieste, Trieste, Italy }
\author[XFEL,FRES]{M. Porro}
\address[FRES]{Department of Molecular Sciences and Nanosystems, Ca’ Foscari University of Venice, Italy}
\author[ALBA]{M. Quispe}
\author[PSI]{B. Schmitt}
\address[PSI]{The Paul Scherrer Institute PSI, Villigen PSI, Switzerland}
\author[DLS]{S. Scully}
\author[XFEL]{M. Turcato}
\author[MAX]{C. Ward}
\author[DESY]{E. Welter}
\address[tri]{The National Institute for Nuclear Physics (INFN), Trieste Section, Italy}
\address[swe]{Department of Computer and Electrical Engineering, Mid Sweden University, Sundsvall, Sweden}




\begin{abstract}
The \textsc{XAFS-DET} work package of the European \textsc{LEAPS-INNOV} project is developing a high-purity Germanium detectors for synchrotron applications requiring spectroscopic-grade response. The detectors integrate three key features: (1) newly designed monolithic Germanium sensors optimised to mitigate charge-sharing events, (2) an improved cooling and mechanical design structure supported by thermal simulations, and (3) complete electronic chain featuring a low-noise \textsc{CMOS} technology based preamplifier, enabling high X-ray count rate capability over a broad energy range (5-100~keV).
This paper discusses the first integration and characterization of one of the two multi-element Ge detectors at the European Synchrotron Radiation Facility (\textsc{ESRF}). The integration phase included validating high-throughput front-end electronics, integrating them with the Ge sensor, and operating them at liquid nitrogen temperature, in addition to the experimental characterization, which consists of electronics noise
study and spectroscopic performance evaluation.
\end{abstract}

\begin{keyword}
\textit{High Purity Germanium sensor, Spectroscopy detectors, XAFS, XRF}
\end{keyword}
\end{frontmatter}
\section{Introduction}
\begin{figure*}
    \centering
    \includegraphics[width=1\linewidth]{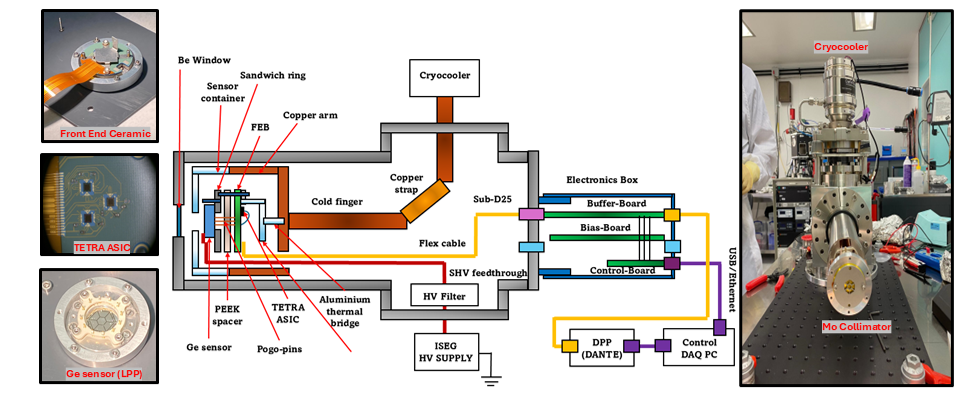}
    \caption{Schematic overview of the XAFS-DET detector setup, illustrating the Ge sensor integration with the front-end and back-end electronics, cryocooler, and data acquisition (DAQ) architecture. The Ge sensor is connected to the TETRA ASIC-based Front-End ceramics \textsc{(FEC)}, which interfaces with the Back-End Board \textsc{(BEB)} for signal amplification and buffering. The \textsc{DANTE} Digital Pulse Processor (DPP) was used instead of Xspress4 for the first integration tests to evaluate electronic noise and detector response. The setup was operated at a temperature of 77~K.}
    \label{fig:Full_schmea}
\end{figure*}

X-ray spectroscopy at synchrotron radiation facilities demands high-performance detectors to sustain high photon flux while maintaining excellent energy resolution. While silicon-based detectors are widely used due to their high resolution capabilities, their limited stopping power restricts their efficiency in the hard X-ray regime. To overcome this limitation, the XAFS-DET subproject~\cite{Orsini2023} is focused on the development of monolithic multi-element Germanium detectors for X-ray Absorption Fine Structure \textsc{(XAFS)}~\cite{calvin2013xafs}, and X-ray Fluorescence Spectroscopy \textsc{(XRF)} applications.
The objective is to improve the sensor's performance by using a monolithic multi-element configuration. This configuration enables a uniform distribution of the X-ray photon flux across the detector sensing elements, ensuring high-resolution detection over a broad X-ray energy range.
In addition, we have also developed a full simulation chain 
 to optimize the detector’s design~\cite{MANZANILLAS2023167904} and thoroughly study the detector response function. 
The full detection setup is showcased using Fig.~\ref{fig:Full_schmea}.
The first integration phase involved rigorous functional testing of the electronic chain, the assembly of the Ge sensor with the front-end electronics, and the first spectroscopic measurements using standard offline X-ray calibration sources. These steps form the basis for further optimization and beamline implementation in future experimental campaigns.
\section{Detector Systems and Experimental Setup}
The detector systems consists of two pixel configurations: (1) a Small Pixel Prototype (SPP) with a 5~mm$^2$ pixel size, and (2) a Large Pixel Prototype (LPP) with a 20~mm$^2$ pixel size. Both sensor architectures incorporate seven active pixels, the centre one being a hexagonal shape, surrounded by six trapezoids and three ancillary pixels sharing their detected events with the guard ring, the latter are designed to reject charge-sharing events to improve spectral performance. The electronics and readout architecture are designed and optimized for an energy range of 5~keV to 100~keV, with the capability of handling count rates between 20~kcps/mm$^2$ to 250~kcps/mm$^2$. It is further connected to the Xpress4~\cite{Xspress4} digital pulse processor (DPP), commercialized by \textsc{quantum detectors}, which helps enhancing the signal-to-noise ratio (SNR) by electronically mitigating the charge-sharing effects, and disentangling almost coincident photons. In addition, the detector cryostat based on an electrical cryocooler \textsc{(ct model)}, provided by \textsc{sun power}, will ensure the detector's stable operation at low temperatures ($\sim$77~K) to minimize noise contribution. At this temperature, the thermally generated dark current is exponentially suppressed, reducing the leakage current and enhancing the charge collection efficiency~\cite{Quispe:2024wis}. 
\subsection{Electronics Tests}
 The electronics of the detection system consists of a multi-channel integrated preamplifier \textsc{(tetra)} incorporated within the Front-End Ceramic (FEC) PCB. Each PCB includes three ASIC preamplifiers, each handling 3-4 readout channels. TETRA, an evolution of \textsc{cube}~\cite{6154396}, is a low-noise CMOS amplifier developed by \textsc{xglab, bruker}. The FEC is then connected to the Back-End Board (BEB), which consists of three subsystems: the Buffer Board, responsible for signal buffering and minimizing distortions before transmission; the Bias Board, which provides power supply regulation, \textsc{ASIC} voltage shifting, and detector temperature monitoring; and the Control Board, which includes an FPGA that manages \textsc{ASIC} operations and overall \textsc{BEB} control. 

\begin{figure}[htb!]
    \centering
\includegraphics[width=1\linewidth]{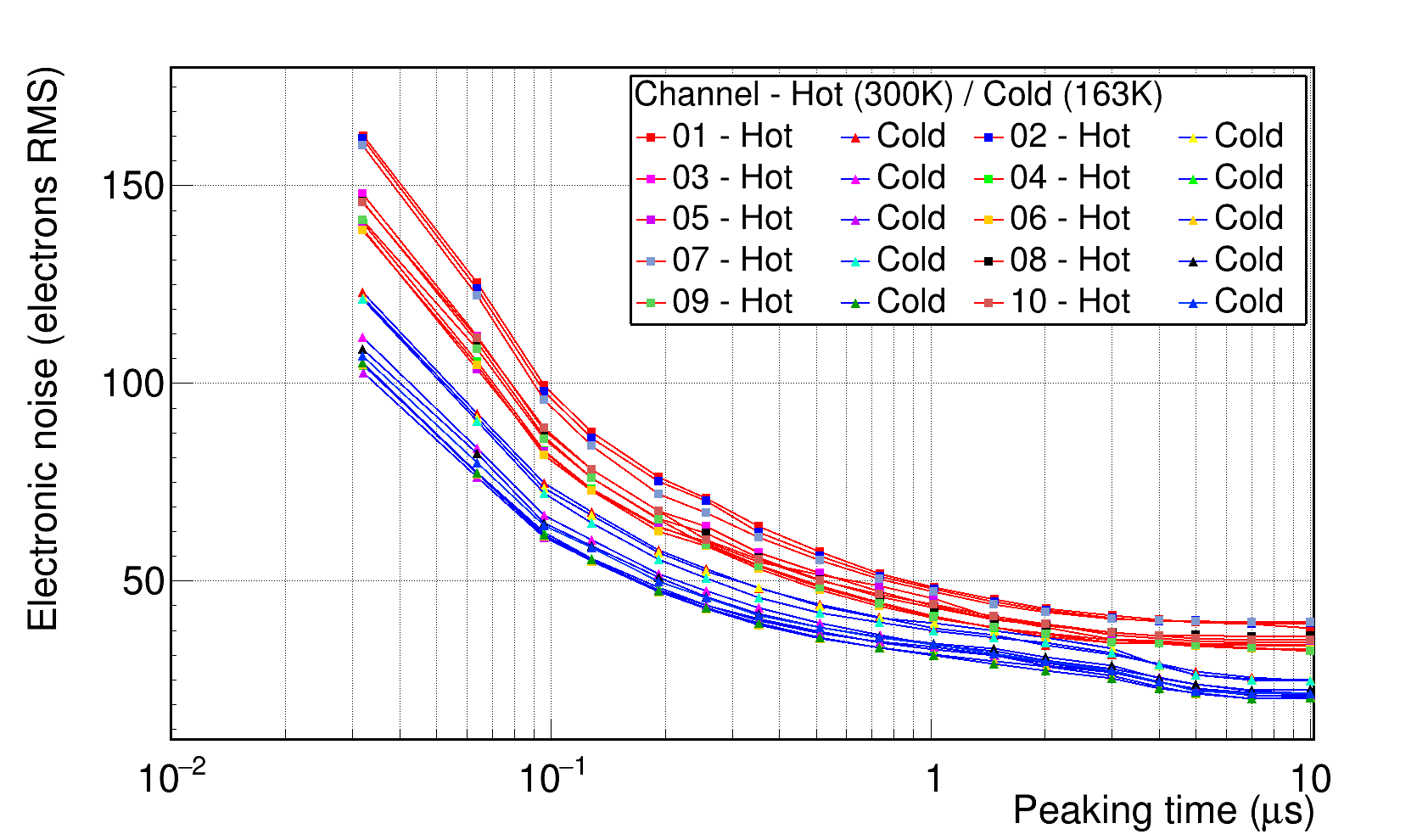}
    \caption{Electronic noise (RMS electrons) as a function of peaking time for 10 channels, measured at room temperature (300~K, 'Hot') and liquid nitrogen temperature (163~K, 'Cold'). The noise decreases significantly at lower temperatures, demonstrating the expected thermal noise reduction in cryogenic conditions.}
    \label{fig:ENCvs.PT}
\end{figure}

 A visual inspection of FEC boards confirmed the proper assembly of ASICs via wire bonding to the front end PCBs, meeting the specifications. Preliminary noise measurements without the Ge sensor were conducted using \textsc{xglab dante} DPP~\cite{Iguaz:2023otz},followed by the characterization of electronics transfer function under three gain configurations, confirming the expected feedback capacitance variation from 30~fF to 174~fF. 
 After the preliminary checks, the electronics were tested again at lower temperatures to assess the noise behaviour. Furthermore, the electronics response at lower temperature (163.15~K) revealed a 10$\%$ reduction in fall time, the time it takes for the signal to decay from its peak value back to baseline, reflecting how quickly the collected charge is processed.
  A notable reduction in the Equivalent Noise Charge (ENC) was observed, decreasing from 50 electrons RMS at room temperature to 30 electrons RMS at (163.15~K) at a peaking time value of 1~µsec (see Fig.~\ref{fig:ENCvs.PT}). Further noise reduction was achieved by installing a dedicated high-voltage RC filter, with a low-pass cutoff frequency of 6.6~Hz. This filter was designed to suppress low-frequency noise components and voltage fluctuations from the high-voltage bias supply.
\subsection{First Integration with a Ge sensor (LPP)}
The LPP Ge sensor was successfully integrated with the FEC  board by aligning the pogo pins to ensure stable electrical connections to the preamplifiers. A thin micron layer of indium foil between the pogo pins and the pixel surfaces was also introduced. We used a 3~mm-thick Molybdenum collimator, which was positioned before the detector, to enhance the \textsc{SNR}. The system was then vacuum-sealed and cooled down to 113.15~K. A bias voltage of +100~V was then applied to the sensor n+ side, resulting in the collection of holes, and the signal was measured on the segmented p+ side. A measured leakage current of less than 0.05~pA, and a reset period of 1~s was recorded with waveform analysis across all 7 inner pixels. With the system fully assembled, the first X-ray energy spectra were acquired, validating the successful integration of the setup.

\section{First Spectroscopic results}
\begin{figure}[htb!]
    \centering
    \includegraphics[width=1\linewidth]{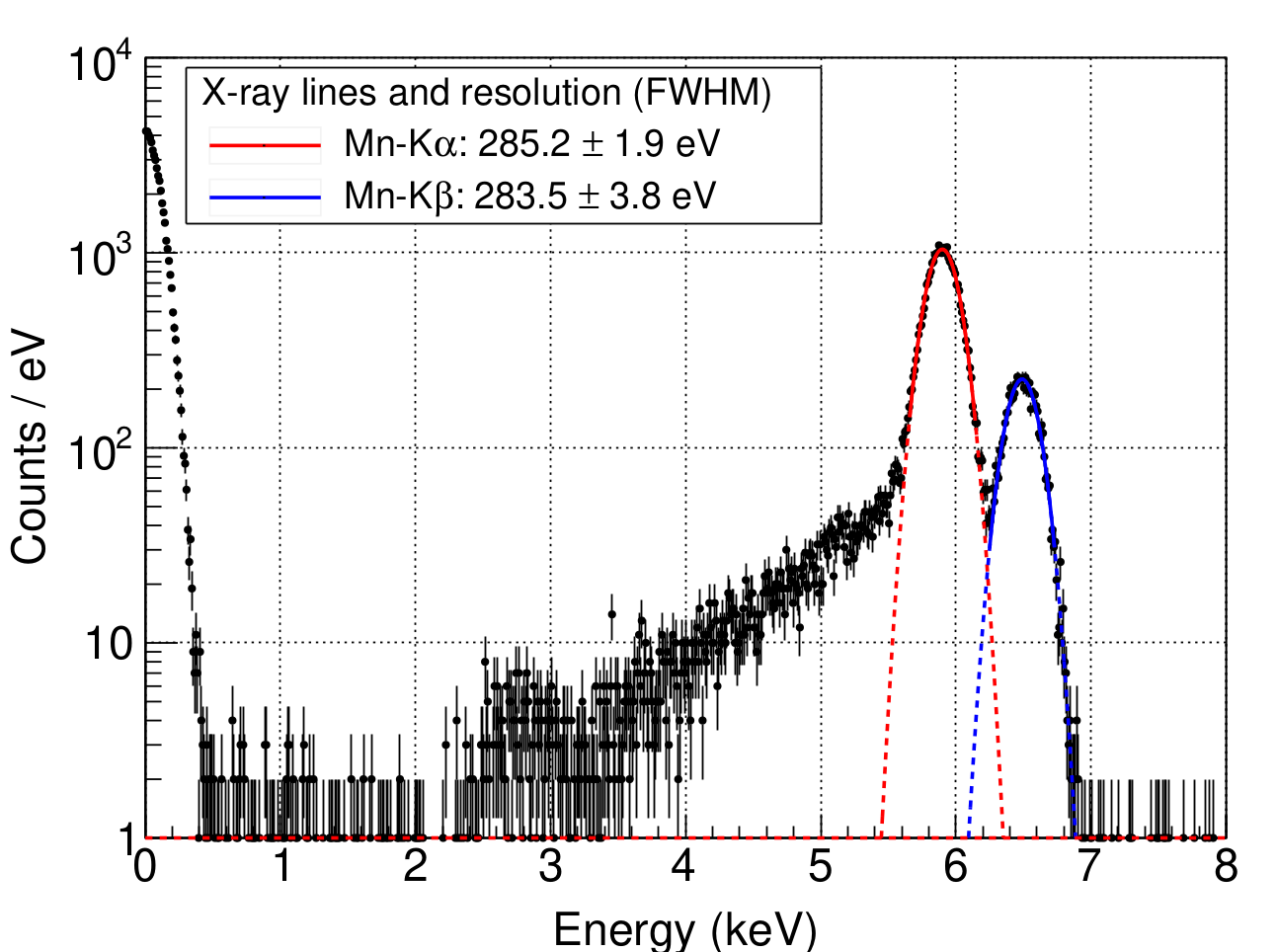}
    \caption{Measured energy spectrum of  $^{55}$Fe radioactive source. The measured energy resolution at Mn-K$\alpha$ at 5.9~keV and Mn-K$\beta$ at 6.4~keV is 285.2$\pm$1.8 and 283.5$\pm$3.8~eV FWHM, respectively. The signal-to-noise ratio \textsc{(SNR)} in this case is defined as the ratio between the fitted peak amplitude of (Mn-K$\alpha$) and the standard deviation of the background in a peak-free region ($\sim$1–3~keV).}
    \label{fig:Mn55}
\end{figure}

\begin{figure}[htb!]
    \centering
    \includegraphics[width=1\linewidth]{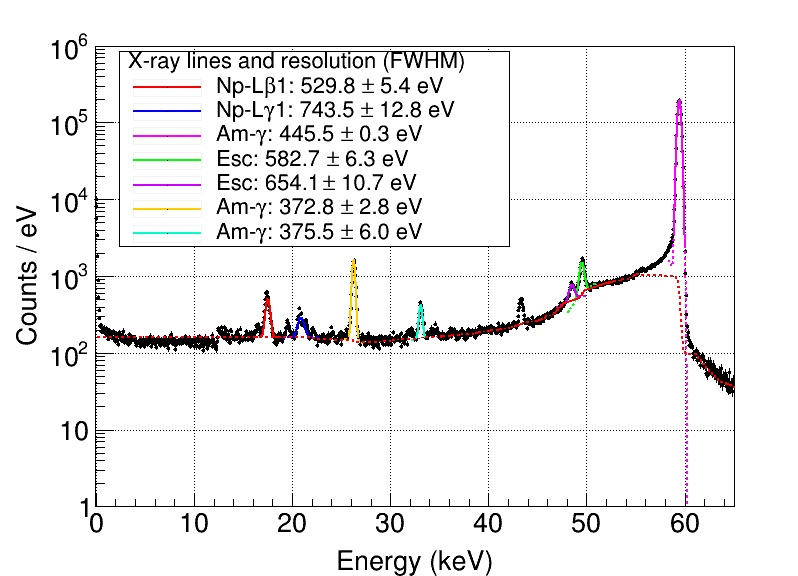}
    \caption{Measured energy spectrum of $^{241}$Am showing characteristic X-ray and gamma-ray lines. The legend highlights energy resolution at the respective peak energy which include: Np-L$\gamma_1$ at ~17.75~keV (red), Np-L$\gamma_2$ at ~20.78~keV (blue), Am-$\gamma$ at ~59.65~keV (Magenta), two Ge escape peaks at ~49.65~keV (Green) and ~48.54~keV (Purple), and two additional $\gamma$-ray contributions at ~26.34~keV (Yellow), and ~33.20~keV (Cyan).}
    \label{fig:AM-241}
\end{figure}

The first spectroscopic tests of \textsc{XAFS-DET} detector were performed using two offline radioactive X-ray calibration sources, namely $^{55}$Fe (Iron-55) and $^{241}$Am (Americium-241), as illustrated in the energy spectra using Fig.~\ref{fig:Mn55} \&~\ref{fig:AM-241}. These sources were chosen due to their well-defined emission lines covering different energy ranges, making them ideal for assessing the detector's response, energy resolution, and stability from low-energy to high-energy X-ray regimes.

The $^{55}$Fe source emits characteristic Mn-K$\alpha$ (5.9~keV) and Mn-K$\beta$ (6.4~keV) X-rays, which are commonly used for energy calibration in the tender X-ray range. In our measurements, the best measurement of FWHM for the Mn-K$\alpha$ peak is 285.2$\pm$1.8~eV at an Input count Rate (ICR) of 3~kcps.
On the higher-energy end, the $^{241}$Am source was used which decays to $^{237}$Np via $\alpha$ emission and gives a prominent gamma-ray emissions at 59.5~keV and several characteristic L X ray lines from its daughter nuclei. The best resolution for the 59.5~keV $\gamma$ peak was measured 445.1$\pm$0.3~eV for an \textsc{ICR} of 2~kcps. 
The detector was also tested with Xspress4 DPP using the $^{55}$Fe source, confirming the energy resolution values. These measurements confirm the functionality and successful integration of the Ge sensor with its associated electronics. 
\begin{figure}[htb!]
    \centering
    \begin{subfigure}[b]{1\linewidth}
        \centering
        \includegraphics[width=1\linewidth]{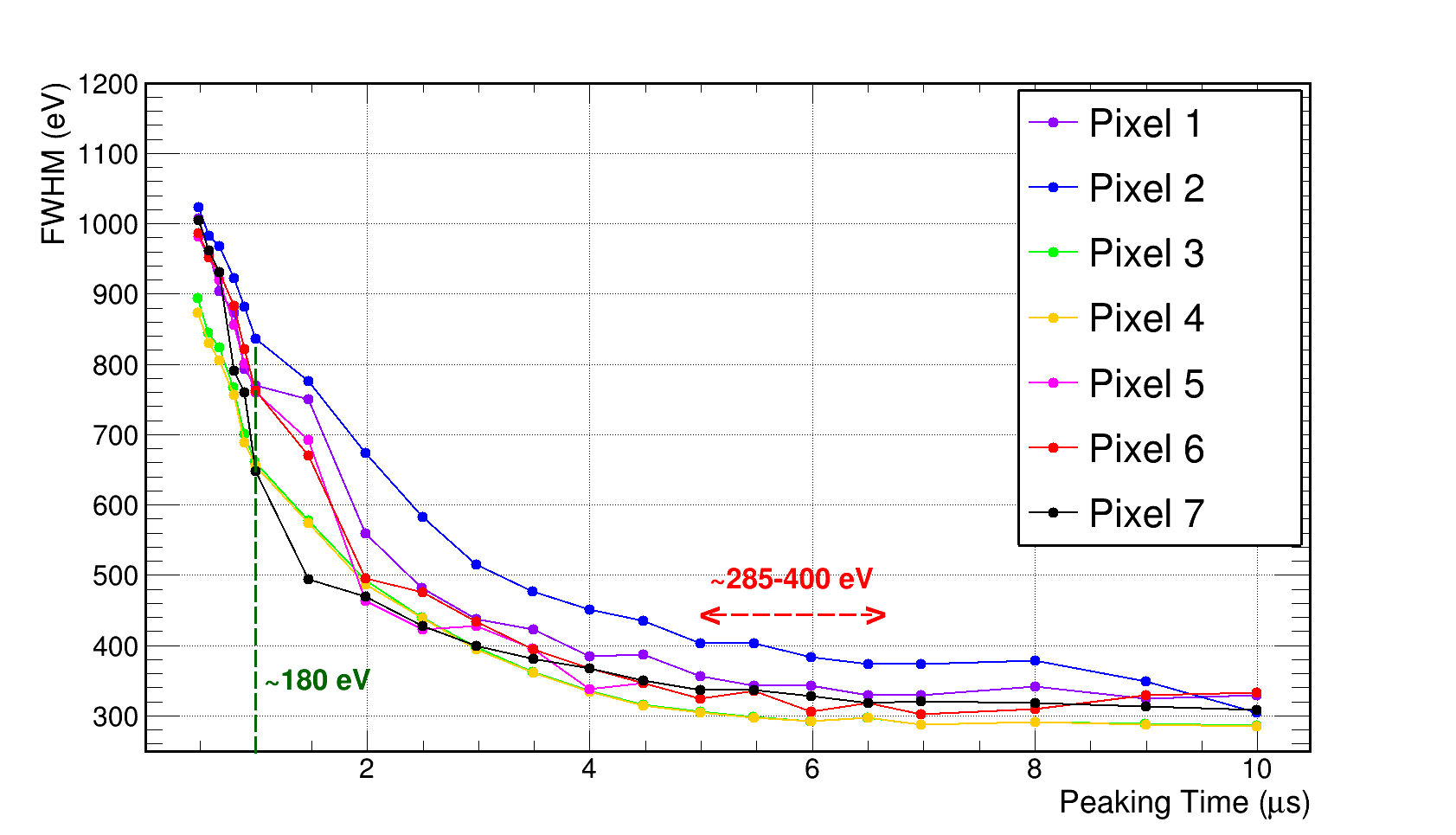}
        \caption{}
        \label{fig:fwhm_lowE}
    \end{subfigure}
    \hfill
    \begin{subfigure}[b]{1\linewidth}
        \centering       
        \includegraphics[width=1\linewidth]{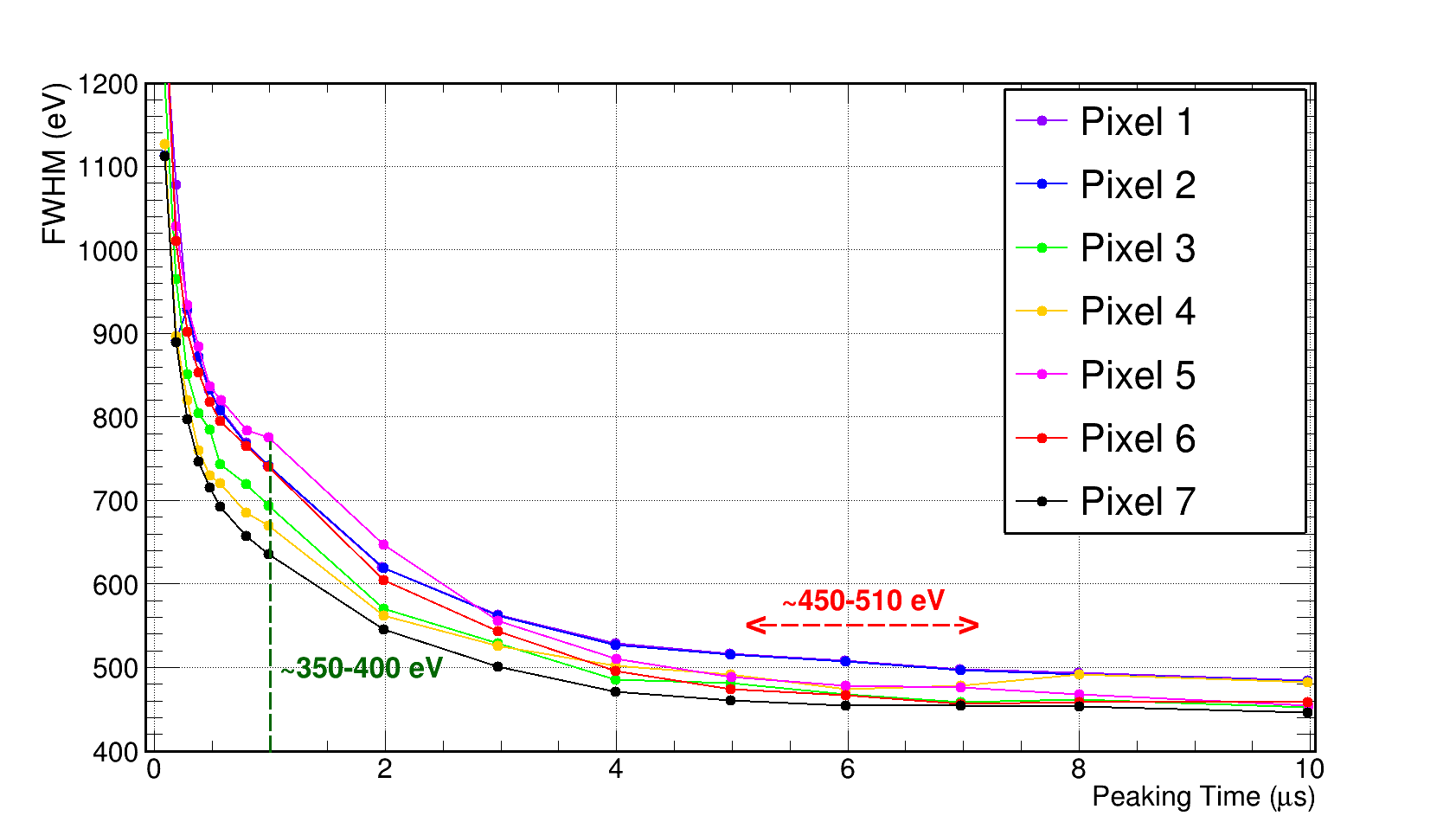}
        \caption{}
        \label{fig:fwhm_highE}
    \end{subfigure}
    \caption{Comparison of energy resolution as a function of PT for (a) low-energy calibration using $^{55}$Fe (5.9~keV) and (b) High-energy calibration using A$^{241}$Am (59.5~keV). The  target resolution at a peaking time value of 1~µs is indicated in green and the best current value at $\sim$6~$\mu$s is indicated in red in both cases.}
    \label{fig:fwhm_comparison}
\end{figure}

A critical parameter affecting energy resolution is the detector’s peaking time, which was optimized to approximately
6.97~$\mu$s, with minimum gap time of 250~ns providing the best overall resolution for this detection setup. The behavior observed in the Fig.~\ref{fig:fwhm_comparison}
matches the theoretical noise response of semiconductor detectors, where a combination of series noise, which is proportional to the detector capacitance, and the noise contribution from the \textsc{ASIC} preamplifiers, parallel noise arising from detector leakage current, and flicker noise contributions coming from the electronics drifts influences the resultant energy resolution~\cite{Spieler:1010490}. At shorter peaking time(1~$\mu$s), series noise dominates, leading to higher ENC and degraded resolution, whereas, the higher values (6-10~$\mu$s) allow for better charge integration, reduced ENC and improved resolution.
However, at very long peaking times, resolution stabilizes as parallel noise and low-frequency noise start dominating. As a goal, our specification is an energy resolution of 180~eV for 5.9~keV energy peak at a peaking time value of 1~$\mu$s to maintain an optimum resolution for the ICR dynamic range up to 1~Mcps.



\section{Current and future developments}
The first integration and tests at ESRF have shown functional detector operation, but there are a few areas to work on to improve the performance.
\begin{enumerate}
    \item 
    Noise Mitigation \& Shielding:
       Imminent ideas include shielding the Back-end Bias Board using a copper-coated enclosure to reduce the electromagnetic interference from the power electronics.
       Other shielding configurations are still to be tested .
    \item 
    Spectroscopy performance improvements: Evaluation of the different contributions to the detector capacitance:  Ge sensor, indium foil, pogo-pins, and the ASICs path.
    \item 
    Prototype Optimisation \& Parallel Testing:
        Further studies will be conducted in a laboratory setting with an X-ray generator before moving to synchrotron testing. A second prototype (Proto II) will be installed to cross-validate the performance.
        \end{enumerate}

\section{Conclusion}
Although initial tests revealed a significant noise contribution, efforts made with HV filtering and particularly in isolating the BEB, have shown refinement in the detector's performance such as the achieved resolution of 285.2$\pm$1.8~eV, and 445.1$\pm$0.3~eV at 5.9~keV and 59.65~keV at 6.97~$\mu$s PT, respectively. In the future, we plan to focus on noise reduction strategies, electromagnetic shielding, spectroscopy performance improvements, and further mechanical optimizations. In addition, we will study how the energy resolution is affected by the photon's impact position on the detector's entrance surface by doing 2D scans with microbeam. The detector will undergo additional beamline validation at the \textsc{ESRF} facility, exploring its performance under high-intensity synchrotron radiation conditions. This step will ensure the detector's readiness for synchrotrons, addressing the needs of spectroscopy applications in a high-photon flux environment while maintaining excellent energy resolution and stability.

\section*{Acknowledgement}
The authors thank the European Union Horizon 2020 programme (grant 101004728) for funding. We are grateful to XGLab S.r.l. team for designing and developing the electronics. We also thank Chouaib Meraihia from ESRF and their detector unit for contributing to electronics testing and sensor integration, Claude Menneglier (SOLEIL) for his assistance in designing mechanical components, and Martin Chauvin (SOLEIL) for his valuable guidance and support.


\bibliography{mybibfile}

\begin{thebibliography}{1}
\expandafter\ifx\csname url\endcsname\relax
  \def\url#1{\texttt{#1}}\fi
\expandafter\ifx\csname urlprefix\endcsname\relax\def\urlprefix{URL }\fi
\expandafter\ifx\csname href\endcsname\relax
  \def\href#1#2{#2} \def\path#1{#1}\fi

\bibitem{Orsini2023}
F.~Orsini, et~al., Xafs-det: A new high throughout x-ray spectroscopy detector system developed for synchrotron applications, Nuclear Instruments and Methods in Physics Research, Section A: Accelerators, Spectrometers, Detectors and Associated Equipment 1045.
\newblock \href {http://dx.doi.org/10.1016/j.nima.2022.167600} {\path{doi:10.1016/j.nima.2022.167600}}.

\bibitem{calvin2013xafs}
S.~Calvin, {XAFS for Everyone}, CRC Press, 2013.
\newblock \href {http://dx.doi.org/10.1201/b14843} {\path{doi:10.1201/b14843}}.

\bibitem{MANZANILLAS2023167904}
L.~Manzanillas, et~al., Development of multi-element monolithic germanium detectors for x-ray detection at synchrotron facilities, Nuclear Instruments and Methods in Physics Research Section A: Accelerators, Spectrometers, Detectors and Associated Equipment 1047 (2023) 167904.
\newblock \href {http://dx.doi.org/https://doi.org/10.1016/j.nima.2022.167904} {\path{doi:https://doi.org/10.1016/j.nima.2022.167904}}.

\bibitem{Xspress4}
G.~Dennis, et~al., {First results using the new DLS Xspress4 digital pulse processor with monolithic segmented HPGe detectors on XAS beamlines}, AIP Conference Proceedings 2054~(1) (2019) 060065.
\newblock \href {http://dx.doi.org/10.1063/1.5084696} {\path{doi:10.1063/1.5084696}}.

\bibitem{Quispe:2024wis}
M.~Quispe, et~al., {Thermal and vibrational studies of a new germanium detector for X-ray spectroscopy applications at synchrotron facilities}, JACoW IPAC2024 (2024) TUPR75.
\newblock \href {http://dx.doi.org/10.18429/JACoW-IPAC2024-TUPR75} {\path{doi:10.18429/JACoW-IPAC2024-TUPR75}}.

\bibitem{6154396}
L.~Bombelli, et~al., {"CUBE", A low-noise CMOS preamplifier as alternative to JFET front-end for high-count rate spectroscopy} (2011) 1972--1975\href {http://dx.doi.org/10.1109/NSSMIC.2011.6154396} {\path{doi:10.1109/NSSMIC.2011.6154396}}.

\bibitem{Iguaz:2023otz}
F.~J. Iguaz, et~al., {DANTE Digital Pulse Processor for XRF and XAS experiments}, JINST 18~(06) (2023) T06011.
\newblock \href {http://arxiv.org/abs/2303.09479} {\path{arXiv:2303.09479}}, \href {http://dx.doi.org/10.1088/1748-0221/18/06/T06011} {\path{doi:10.1088/1748-0221/18/06/T06011}}.

\bibitem{Spieler:1010490}
H.~Spieler, \href{https://cds.cern.ch/record/1010490}{{Semiconductor detector systems}}, Series on semiconductor science and technology, Oxford Univ. Press, Oxford, 2005.
\newblock \href {http://dx.doi.org/10.1093/acprof:oso/9780198527848.001.0001} {\path{doi:10.1093/acprof:oso/9780198527848.001.0001}}.
\newline\urlprefix\url{https://cds.cern.ch/record/1010490}

\end{thebibliography}
\bibliographystyle{model4-names.bst}
\end{document}